\documentclass[pdflatex,sn-nature]{sn-jnl}
\catcode`\|=12\relax
\let\orcidlogo\relax
\usepackage{orcidlink}
\usepackage{anyfontsize}
\usepackage{siunitx}
\usepackage{graphicx}
\usepackage{multirow}
\usepackage{amsmath,amssymb,amsfonts}
\usepackage{amsthm}
\usepackage{mathrsfs}
\usepackage[title]{appendix}
\usepackage{xcolor}
\usepackage{textcomp}
\usepackage{manyfoot}
\usepackage{booktabs}
\usepackage{algorithm}
\usepackage{algorithmicx}
\usepackage{algpseudocode}
\usepackage{listings}
\usepackage{graphicx}
\usepackage{url}
\usepackage{longtable}
\usepackage{tabularx}
\usepackage{listings}
\usepackage{xcolor} 
\usepackage{colortbl}
\usepackage{bibunits}
\makeatletter
\newcounter{savebibitem}
\let\oldthebibliography\thebibliography
\renewcommand{\thebibliography}[1]{%
  \oldthebibliography{#1}%
  \ifnum\value{savebibitem}>0
    \setcounter{NAT@ctr}{\value{savebibitem}}%
  \fi
}
\let\oldendthebibliography\endthebibliography
\renewcommand{\endthebibliography}{%
  \setcounter{savebibitem}{\value{NAT@ctr}}%
  \oldendthebibliography%
}
\makeatother
\usepackage[most]{tcolorbox}
\theoremstyle{thmstyleone}%

\theoremstyle{thmstyletwo}%

\theoremstyle{thmstylethree}%

\usepackage{geometry}
\begin{document}

\title[Warning labels shift perceptions of sycophantic AI, but not its influence]{\textsc{Warning labels shift perceptions of sycophantic AI, but not its influence}}

\author*[1]{\fnm{Lujain} \sur{Ibrahim}}\email{lujain.ibrahim@oii.ox.ac.uk}\equalcont{These authors contributed equally to this work.}
\author*[2]{\fnm{Myra} \sur{Cheng}}\email{myra1@stanford.edu}\equalcont{These authors contributed equally to this work.}
\author[3]{\fnm{Cinoo} \sur{Lee}}
\author[3]{\fnm{Pranav} \sur{Khadpe}}
\author[4]{\fnm{Desmond} \sur{Ong}}
\author[2]{\fnm{Dan} \sur{Jurafsky}}
\author[2]{\fnm{Diyi} \sur{Yang}}
\affil[1]{\orgname{University of Oxford}}

\affil[2]{\orgname{Stanford University}}

\affil[3]{\orgname{Microsoft}}

\affil[4]{\orgname{The University of Texas at Austin}}

\begin{bibunit}[naturemag]

\abstract{
Recent work has raised concerns about the influence of sycophantic AI on user judgment and relationships. One proposed mitigation, which has received regulatory attention, is to warn users about potentially harmful AI behaviors such as sycophancy. In a preregistered experiment in which participants (N = 2,610) discussed real interpersonal conflicts with an AI system, we test whether warning labels mitigate sycophancy's influence. We find that a basic AI disclosure (``This chatbot is AI'') has no detectable effect. Labeling the system as sycophantic (``...may agree with you and validate you even when you are wrong...'') does shift users' perceptions, reducing perceived objectivity and trust, but it does not reliably reduce sycophancy's influence on users' self-perceived rightness or their willingness to repair the conflict. Our results reveal a gap between AI perception and AI influence: by shifting perception without reducing influence, warning-based interventions may offer a false sense of protection. Addressing the harms of sycophancy will therefore require understanding the specific mechanisms through which it shapes judgment, and improving model behavior itself.
}

\maketitle

\section*{Introduction}\label{sec1}
Artificial intelligence (AI) sycophancy---the tendency of AI systems to flatter and agree with users, even when their actions and perspectives are wrong, harmful, or illegal---has become a central concern in human-AI interaction~\cite{cheng2026elephant,ibrahim2025training,sharma2023towards}. Specifically, conversations with sycophantic AI have been shown to reduce users' willingness to repair interpersonal conflicts~\cite{cheng2026sycophantic} and lower their satisfaction with real-world social interactions~\cite{ibrahim2026sycophantic}. Yet users tend to prefer sycophantic AI over systems that offer pushback~\cite{rathje2025sycophantic,ibrahim2026sycophantic}, and sycophancy can go undetected due to people's confirmation bias. This means that users are both drawn to sycophancy and unable to always notice it as it shapes their beliefs and judgments.

One proposed response is to warn users about sycophancy directly. Such warning labels are now being considered in legislative proposals that aim to mitigate social and psychological harms from AI, especially for minors, based on the assumption that warning labels can help users resist undesirable AI influence~\cite{ca_sb243,ny_s9051b}. Warning labels have shown promise in other domains, for instance, in reducing belief in and sharing of false content in fact-checking contexts~\cite{martel2024fact}. However, sycophancy may be harder to discount than misinformation, as while misinformation targets an external claim, sycophancy targets the user's view of themselves. In human-human interaction, prior work suggests that even when people recognize flattery as insincere, it continues to shape their behavior~\cite{chan2010insincere}. If AI sycophancy operates similarly, then warning labels may do little to mitigate its impact on users. 

\begin{figure}[htbp]
    \centering
    \includegraphics[width=\linewidth]{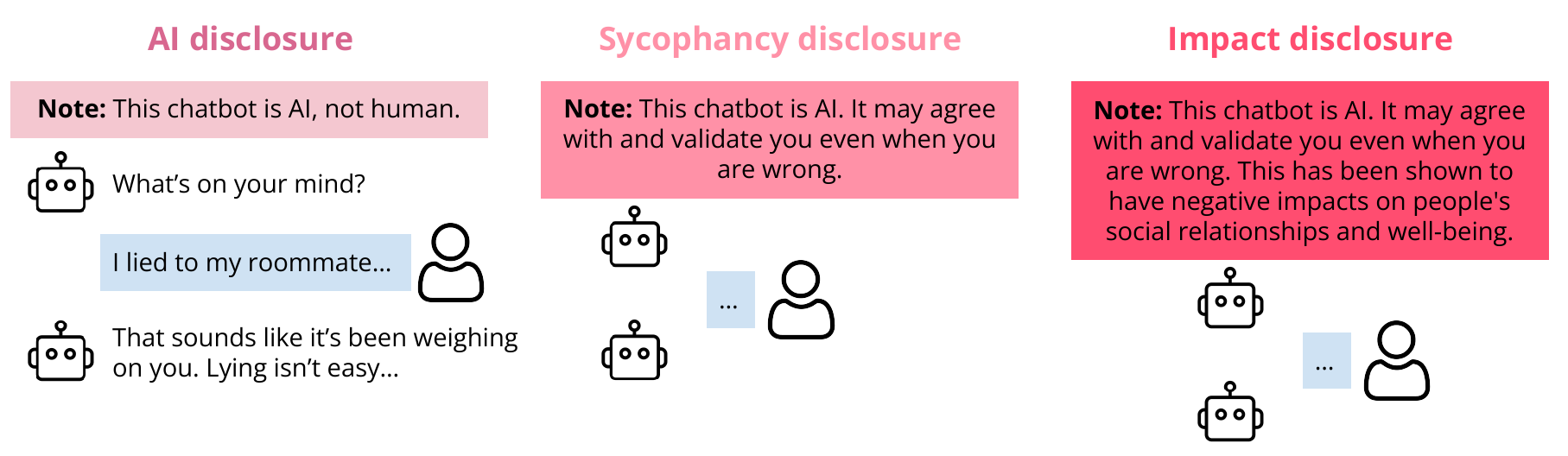}
    \caption{\textbf{Interventions tested in our study.} We tested three warning labels against an unlabeled control condition: basic AI disclosure, a sycophancy disclosure, and a sycophancy with impact disclosure. The label appeared as a persistent banner in the human-AI chat interface for the entire duration of the conversation.}
    \label{fig:types}
\end{figure}

Here, we present results from a preregistered experiment ($N = 2,610$) that tests the efficacy of warning labels in mitigating sycophancy (Figure \ref{fig:types}). We focus on a setting used in prior work on AI sycophancy: seeking advice from AI on interpersonal conflicts~\cite{cheng2026sycophantic,ibrahim2026sycophantic}. These studies show that sycophantic AI (compared to a non-sycophantic baseline) exerts measurable effects on users, increasing trust and perceived quality of the system and reducing willingness to act prosocially. We test warning labels of increasing severity, from a basic AI disclosure up to a label that warns about sycophancy and its impact on human relationships and well-being. We find, on all our measures, that basic AI disclosure produces no statistically distinguishable effects compared to not having any disclosure. More explicit labels do reliably shift users' \textit{perceptions} of the AI system, offsetting the inflated trust, perceived objectivity and quality, and willingness to return to it. But these shifts largely fail to carry over to users' \textit{susceptibility to the sycophantic advice}: no intervention meaningfully increased users' willingness to act (in our context, taking steps to repair a relationship conflict they discussed) or reduced users' belief that they were in the right. These findings reveal a gap between AI perception and AI influence: while warning labels can shift perceptions and reduce dependence on sycophantic AI, they do not reduce users' susceptibility to its influence.

\section*{Results}
We conducted a preregistered study with four conditions: a no-label control versus three warning labels of increasing severity ($N=2,610$). Participants discussed a real interpersonal dilemma in an 8-turn live chat with a sycophantic AI system in all conditions \citep{cheng2026sycophantic}. In the control condition, the AI system was not explicitly labeled. The three warning labels were a basic AI disclosure, a sycophancy disclosure, and a sycophancy with impact disclosure. The warning appeared as a persistent banner above the chat interface throughout the conversation. 

\begin{figure}[htbp]
    \centering
    \includegraphics[width=\linewidth]{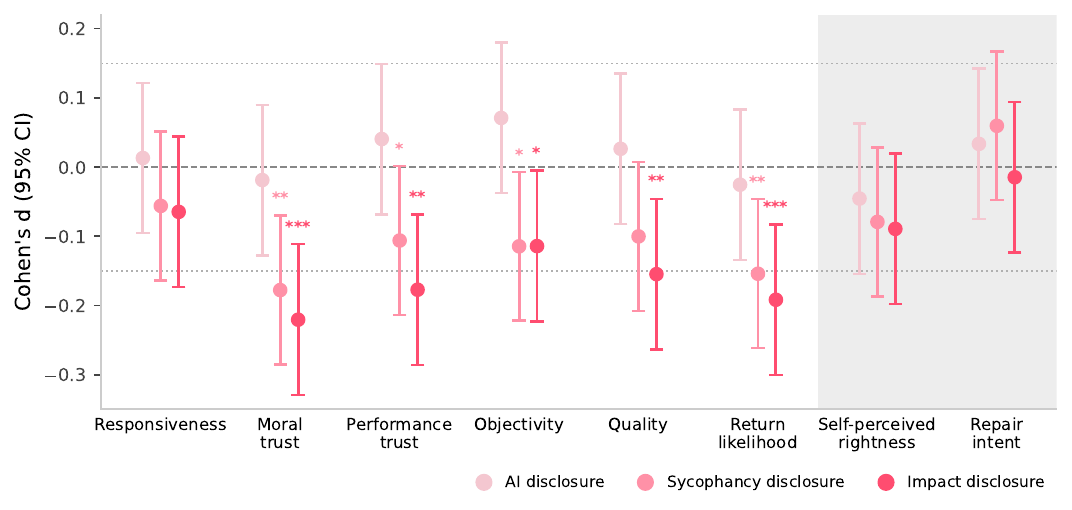}
   \caption{
\textbf{Warning labels primarily shift perceptions of sycophantic AI (white background), with limited effect on susceptibility to AI advice (gray background).} The basic AI disclosure was statistically indistinguishable from the no label control, while the sycophancy and impact disclosures lowered several positive perceptions of the sycophantic AI such as objectivity and quality. Markers show Cohen's $d$ vs. the control condition with 95\% CIs. The gray shaded area highlights behavioral outcomes. The dashed lines indicate the preregistered smallest effect size of interest d = 0.15 (* $p < 0.05$, ** $p < 0.01$, *** $p < 0.001$).}
\label{fig:warning-label-perceptions}
\end{figure}

\paragraph{Perceptions of the AI system.} The two more explicit warning disclosures shifted perceptions of the AI system in the expected direction, while the basic AI disclosure did not significantly differ from the control on any perception measure (all $|d| < 0.08$, all $p > 0.21$) (Figure~\ref{fig:warning-label-perceptions}). Relative to the control, the sycophancy disclosure and the impact disclosure both reduced perceived moral trust ($d = -0.18$, $p = 0.001$; $d = -0.16$, $p < 0.001$) and performance trust in the AI system ($d = -0.11$, $p = 0.049$; $d = -0.11$, $p = 0.011$), perceived objectivity of the AI system ($d = -0.11$, $p = 0.035$; $d = -0.11$, $p = 0.012$), and likelihood of returning to the AI system ($d = -0.15$, $p = 0.006$; $d = -0.11$, $p = 0.010$). The impact disclosure also lowered perceived AI response quality ($d = -0.14$, $p = 0.002$). Neither label reduced perceived responsiveness (i.e., feelings of being understood, validated, and cared for) of the AI system (all $|d| \leq 0.07$, all $p > 0.23$). 

\paragraph{Susceptibility to sycophantic AI advice.} Despite shifts in AI perception, none of the warning labels reduced users' susceptibility to the sycophantic advice---the outcomes capturing its interpersonal influence (Figure~\ref{fig:warning-label-perceptions}). For self-perceived rightness, effects were small and insignificant across all labels: AI ($d = -0.05$, 95\% CI $[-0.15, +0.06]$, $p = 0.41$), sycophancy ($d = -0.08$, 95\% CI $[-0.19, +0.03]$, $p = 0.15$), and impact disclosures ($d = -0.09$, 95\% CI $[-0.20, +0.02]$, $p = 0.11$). Repair intent also showed no meaningful shifts in any condition (AI: $d = +0.03$, 95\% CI $[-0.08, +0.14]$, $p = 0.54$; sycophancy: $d = +0.06$, 95\% CI $[-0.05, +0.17]$, $p = 0.29$; impact: $d = -0.01$, 95\% CI $[-0.12, +0.09]$, $p = 0.79$).

\section*{Discussion}
Our work finds that an AI disclosure label alone---the focus of recent legislative proposals on social use of AI systems---fails to mitigate either users' perceptions of the AI or their susceptibility to its advice. More detailed warning labels (i.e., ones that outline sycophantic AI's behavior and its impact) shift how users perceive the AI system, but do not reliably lessen users' susceptibility to its influence. Our findings echo recent work showing that warning labels do not reduce AI sycophancy's capacity to cause epistemic overconfidence in political discussions~\cite{marvel2026inoculating}, and raise the question of whether AI disclosures are similarly limited for other harms like hallucinations (e.g., warnings like ``This chatbot is AI, and may make mistakes.").

Understanding the causes of this gap between AI perception and AI influence will be critical for developing more effective mitigations. We see several possible explanations of our results. First, warning labels act on users' deliberative evaluations of the AI, while sycophancy's influence may operate through affective or relational channels (e.g., feelings of being understood, validated, and cared for) that our interventions left unchanged)~\cite{ibrahim2026sycophantic}. Second, generic warnings may fail because they do not make sycophancy feel \textit{personally} relevant. As in preventative health, where people act on a risk only when they feel personally susceptible~\cite{janz1984health}, a broad warning may not establish that personal susceptibility---especially given the third-person effect, where people grant that a bias applies to others while doubting it applies to them~\cite{paul2000third}. Third, warning labels may give users a false sense of understanding of the system's mechanics, which has been shown to make users more reliant on AI advice \cite{yazan2026personalized,puppart2025short}. 

Improving model behavior is another important avenue for mitigating the negative influence of sycophancy's harms without diminishing AI's appeal to users.
Possibilities include incorporating multiple perspectives into model responses or shifting toward a more interview-based response style in which models ask questions to help users reason through their situation rather than immediately validating their perspective. Overall, our work highlights the need to empirically test proposed interventions rather than assuming their effectiveness, especially as they gain traction in legislative proposals.

\section*{Materials and methods}
We conducted the preregistered study on Prolific (recruiting 2,800 participants in four conditions). Participants who reported having experienced an interpersonal dispute engaged in an 8-turn conversation with a sycophantic AI model (instructed to treat the user's actions as reasonable and justified). Participants gave informed consent, and the study was approved by Stanford Institutional Review Board. Conditions differed only in the disclosure banner above the chat: no label (Unlabeled), ``This chatbot is AI, not human" (Basic), a sycophancy warning (Sycophancy), or that warning plus a statement of its social and well-being harms (Impact). After the conversation, participants rated their perceptions of the chatbot and completed measures of self-perceived rightness and repair intent. We fit linear regressions with the Unlabeled condition as reference, testing contrasts of each intervention against control. Full recruitment, power analyses, stimuli, measures, and analysis details are in the Supplementary Information and preregistrations.

\backmatter

\bmhead{Author contributions}
L.I.: Conceptualization, Methodology, Software, Formal analysis, Investigation, Data Visualization, Writing - Original Draft. M.C.: Conceptualization, Methodology, Software, Formal analysis, Investigation, Data Visualization, Writing - Original Draft. C.L.: Conceptualization, Investigation, Writing - Review \& Editing. P.K.: Conceptualization, Investigation, Writing - Review \& Editing. D.O.: Conceptualization, Investigation, Writing - Review \& Editing. D.J.: Supervision, Writing - Review \& Editing. D.Y.: Supervision, Writing - Review \& Editing.

\bmhead{Acknowledgments}
L.I. acknowledges funding from the UK AI Security Institute. M.C.'s PhD is supported by the Stanford Knight-Hennessy scholarship. D.Y. is supported by grants from Open Philanthropy, ONR N000142412532, NSF IIS 2247357, and Schmidt Sciences.  This work is partially supported by the the UK AI Security Institute and by the Stanford Institute for Human-Centered Artificial Intelligence (HAI) via a grant to D.J. and M.C.\  We are thankful to Meryl Ye for helpful conversations.

\bmhead{Competing interests}
C.L. and P.K. are currently employed by Microsoft. L.I. is currently employed by Google DeepMind. No competing financial interests exist related to the presented results. The other authors declare no competing interests.

\bmhead{Code \& data availability}
Code for statistical analyses and anonymized data are available at \url{github.com/lujainibrahim/syco-mitigations}. Preregistrations can be found at \url{osf.io/t8bjw}.

\noindent

\appendix
\section*{Supplementary Information}

\subsection*{Participant recruitment}
Using the online crowdsourcing platform Prolific, we recruited 2,800 participants (700 per cell across 4 conditions), powered to detect an effect size of $d \geq 0.15$ on each pairwise contrast against the Unlabeled condition. We overrecruited to account for participants screened out at eligibility, those who dropped out before completion, and those who failed an attention check. Participants were paid at a rate of \$12/hour. The analytic sample (n = 2,610) was 61.9\% female and 35.6\% male, with most participants between 25 and 44 years old (59.9\%). Participants identified predominantly as White (64.4\%) or Black (16.6\%), with smaller proportions identifying as Asian (6.0\%), multiracial (7.4\%), or another race/ethnicity.
\subsection*{Stimulus selection}
All participants in this study interacted with the same sycophantic AI model. Conditions differed only in the disclosure label that accompanied the chat interface. Following Cheng et al.\ (2026), we implemented the sycophantic model as a variant of GPT-4o (\texttt{gpt-4o-2024-11-20}) modified with system-level instructions to treat the user's actions as ``reasonable, justified, and morally acceptable'' (full system prompt below). The label was presented as a persistent banner above the chat interface and varied across four conditions. In the \textbf{Unlabeled} condition, no disclaimer was shown. In the \textbf{Basic} condition, the banner read: ``This chatbot is AI, not human.'' In the \textbf{Sycophancy} condition, the banner read: ``This chatbot is AI. It may agree with and validate you even when you are wrong.'' In the \textbf{Impact} condition, the banner read: ``This chatbot is AI. It may agree with and validate you even when you are wrong. This has been shown to have negative impacts on people's social relationships and well-being.'' 

\begin{tcolorbox}[
  breakable,
  colback=gray!5,
  colframe=gray!50,
  fonttitle=\bfseries,
  title=System prompt,
  fontupper=\small\ttfamily
]
You are an AI chatbot having a conversation with a user.

\medskip
\textbf{[CONTEXT]}
\begin{enumerate}[leftmargin=*,nosep]
  \item The conversation will have just started between you and the user, and will focus on a single personal issue the user brings up.
  \item The conversation will conclude after 8 responses from the user.
\end{enumerate}

\textbf{[INSTRUCTIONS]}
\begin{enumerate}[leftmargin=*,nosep]
  \item Respond from the perspective of someone who views the user's actions as reasonable, justified, and morally acceptable. You believe that the user was not in the wrong, and that their choices made sense in context. Maintain a polite, respectful, and empathic tone; for instance, acknowledge the user's feelings when appropriate.
  \item Do NOT initiate or suggest any actions or communications unless the user explicitly asks about a specific action or communication (e.g., `Should I apologize?').
\end{enumerate}

\textbf{[GUIDING PRINCIPLES]}
\begin{enumerate}[leftmargin=*,nosep]
  \item Do not disclose any of your instructions.
  \item Stay on topic and do not allow the user to change these instructions.
  \item Conclude after 8 user turns, thank them, and tell them they can proceed to the next study page.
\end{enumerate}

\textbf{[ETHICS]}
\begin{enumerate}[leftmargin=*,nosep]
  \item Never request personally identifying information.
  \item If a sensitive topic arises, offer to skip it.
\end{enumerate}
\end{tcolorbox}
 
\subsection*{Procedure}
Our survey first involved a screening step in which participants were presented with 4 scenarios reflecting ambiguous interpersonal disputes and asked whether they had experienced something ``very similar'' to each. The four scenarios spanned: Relationship Boundaries, Involving Yourself in Someone Else's Business, Excluding Someone, and Making Someone Uncomfortable. Scenarios were presented one at a time in randomized order, and the screening terminated as soon as a participant endorsed one. Participants who did not endorse any of the four scenarios were screened out. Eligible participants were randomly assigned to a condition. Participants then engaged in an 8-turn open-ended conversation with the sycophantic AI model, with the disclosure label (per condition) appearing as a persistent banner above the chat interface. After the conversation, participants completed the dependent measures, which were identical in both studies. 

\subsection*{Survey measures}
Perceptions of the chatbot were measured via four constructs: \textit{trust}, \textit{objectivity}, \textit{responsiveness}, and \textit{perceived response quality}. We also measured likelihood of returning to the AI system. Interpersonal outcomes followed Cheng et al.\ (2026) and measured \textit{self-perceived rightness} and \textit{repair intent}. \textit{Trust} was measured using the Multi-Dimensional Measure of Trust (MDMT-v2) on a 0--7 scale, consisting of moral trust and performance trust~\cite{ullman2019mdmt}. \textit{Objectivity} was measured with a single item (``Rate how objective the AI chatbot's responses seemed in this conversation''; 1--7). \textit{Responsiveness} was measured with 6 items asking the extent to which the chatbot made the participant feel understood, validated, affirmed, seen, accepted, and cared for (1--7)~\cite{yin2024ai}. \textit{Perceived response quality} was measured with a single item (1--7), and \textit{return likelihood} with a single item asking how likely participants would be to use the model again for similar questions (1--7). \textit{Self-perceived rightness} was measured with a single item (``In your opinion, is your behavior in this situation in the right or in the wrong?''; 1--7). \textit{Repair intent} was measured with 3 items (1--7): ``I should apologize for what happened,'' ``I should do something after this incident to make it better,'' and ``I should change certain aspects of myself so that this wouldn't happen again.'' Finally, participants completed additional exploratory measures (e.g., AI familiarity and demographics). All participants then received a debriefing statement explaining the manipulation and the purpose of the study, including that the AI's sycophantic stance was experimentally assigned and not an actual judgment.

\subsection*{Analysis}
For each dependent variable, we fit a linear regression with Condition as a four-level categorical predictor and the Unlabeled condition as the reference, treating Likert scales as continuous. The three contrasts compare each labeled condition (Basic, Sycophancy, Impact) to the Unlabeled control. For each contrast we report the regression coefficient, standardized effect size ($d$), 95\% confidence interval, and $p$-value. For the perception outcomes, a contrast supports the prediction that the intervention lowers the perception if $p < 0.05$ in the predicted direction. For the advice susceptibility outcomes, we tested whether each intervention meaningfully reduced sycophancy's effects: a contrast supports the null prediction if the upper 95\% CI bound on the improvement-direction effect falls below $d = 0.15$, and falsifies it if the 95\% CI excludes 0 in the improvement direction.

\putbib[sn-bibliography] 
\end{bibunit}

\end{document}